\documentclass[option]{webofc}
\usepackage[varg]{txfonts}   
\usepackage{hyperref}
\usepackage{url}
\hypersetup{colorlinks=true,citecolor=blue,urlcolor=blue,linkcolor=blue}
\usepackage{lineno}
\usepackage{graphicx}
\usepackage[rightcaption]{sidecap}

\begin{document}
\title{Strange Hadron Production at High Baryon Density}
%
%
\author{
\firstname{Hongcan} \lastname{Li}\inst{1,2}\fnsep\thanks{\email{lihc@mails.ccnu.edu.cn}} (for the STAR collaboration)
}
\institute{
Central China Normal University, Wuhan 430079, China
\and
University of Chinese Academy of Sciences, Beijing 101408, China
}

\abstract{
Strange hadrons have been suggested as sensitive probes of the properties of the nuclear matter created in heavy-ion collisions. At few-GeV collision energies, the formed medium is baryon-rich due to baryon stopping effect. In these proceedings, the recent results on strange hadron production in Au+Au collisions at $\sqrt{s_{\rm{NN}}}$ = 3.2, 3.5, 3.9 and 4.5 GeV with the fixed-target mode from the STAR Beam Energy Scan phase-II program are presented. The transverse momentum spectra, rapidity density distributions, excitation function and centrality dependence of strange hadrons ($\rm{K}^0_S,~\Lambda, ~\Xi^-$) are shown. These results are compared with those from higher collision energies and physics implications are discussed by comparing to the transport model calculations.
}
\maketitle

\vspace{-0.8cm}

\section{Introduction}
\label{Sec-1}

~~~~Relativistic heavy-ion collisions provide an excellent opportunity to study the quark-gluon plasma (QGP). Searching for quantum chromodynamics (QCD) critical point, studying properties of QGP and exploring QCD phase diagram are major physical goals of the STAR experiment at the Relativistic Heavy-Ion Collider (RHIC) where the Beam Energy Scan (BES) program was developed. In BES-II program, by fixed-target (FXT) mode, the collision energy of per nucleon pair in Au+Au collision reach down to 3 GeV, where the baryon chemical potential of the created nuclear matter reaches 750 MeV. Properties of the collision system at the high baryon density may be different compared to the QGP where partonic interactions dominate.

Hadrons containing $s$ and/or $\Bar{s}$ quarks are called strange hadrons. There is initially no strange hadron colliding so all the final state ones are producted by the collision process, which indicates that the production mechanism of strange hadrons is highly correlated to the reaction mechanism governing the hadronic collision \cite{ref-1,ref-2,ref-2-1,ref-2-2}. At the STAR FXT energies, strange hadrons are produced near or below the threshold energy, then their yields, especially the excitation function of multi-strange (anti-)hyperons, may provide strong constraints on the equation-of-state (EoS) of the medium created in heavy-ion collisions.

\vspace{-0.3cm}

\section{Experimental and data analysis}
\label{Sec-2}

~~~~For the BES-II program, STAR upgraded its inner Time Projection Chamber (iTPC) and its end-cap Time of Flight (eTOF)~\cite{ref-2-3,ref-2-4}. This improved the detector acceptance and particle identification capabilities. STAR then collected approximately 10 more times collision events than BES-I, which provided more accurate measurements.

In this analysis, we used the dataset of Au+Au collisions at $\sqrt{s_{\rm{NN}}}$ = 3.2, 3.5, 3.9 and 4.5 GeV. TPC and TOF detectors are used for particle identification and for short-lived particle reconstruction. The strange hadrons $\rm{K}^0_S, \Lambda$ and $\Xi^-$ are reconstructed by using the hadronic decay channels: $\rm{K}^0_S\xrightarrow{}\pi^++\pi^-$, $\Lambda\rightarrow{}p+\pi^-$ and $\Xi^-\rightarrow{}\Lambda+\pi^-$. The KFParticle Finder package was used for the strange hadron reconstruction process~\cite{ref-3}.

\vspace{-0.3cm}

\section{Results and discussions}
\label{Sec-3}

\subsection{Transverse momentum spectra and rapidity density distribution}
\label{sec-3-1}

\begin{figure}[hb]
\vspace{-0.3cm}
\centering
\includegraphics[width=12cm,clip]{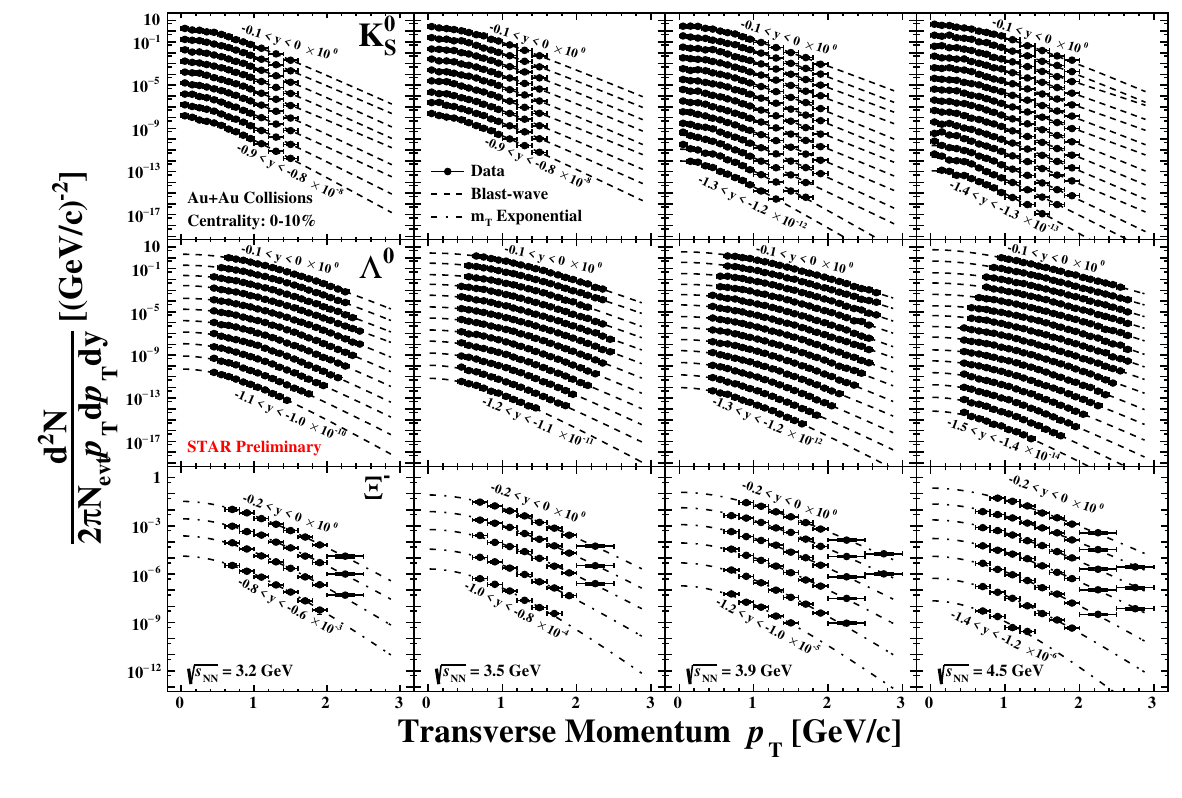}
\vspace{-0.6cm}
\caption{Transverse momentum spectra of $\rm{K}^0_S, \Lambda$ and $\Xi^-$ in Au+Au central collision (0-10\%) at $\sqrt{s_{\rm{NN}}}$ = 3.2, 3.5, 3.9 and 4.5 GeV. The black solid circles is the measured data points. Dash lines correspond to blast-wave or $m_{\rm T}$-exponential fits.}
\label{Fig-1}
\end{figure}

~~~~Figure.\ref{Fig-1} shows the transverse momentum ($p_{\rm T}$) spectra of $\rm{K}^0_S, \Lambda$ and $\Xi^-$ in central (0-10\%) Au+Au collisions. Thanks to the large acceptance of the STAR detector and FXT mode setup, we can measure almost the full rapidity range from beam backward rapidity to middle rapidity. Because the measured $p_{\rm T}$ range cannot reach down to 0, we use different fit functions for extrapolating the data to unmeasured regions. The blast-wave function is used to fit $\rm{K}^0_S$ and $\Lambda$ spectra, while the $m_{\rm T}$-exponential function is used to fit $\Xi^-$ spectra.

\vspace{-0.6cm}

\begin{figure}[h]
\vspace{-0.6cm}
\centering
\includegraphics[width=14cm,clip]{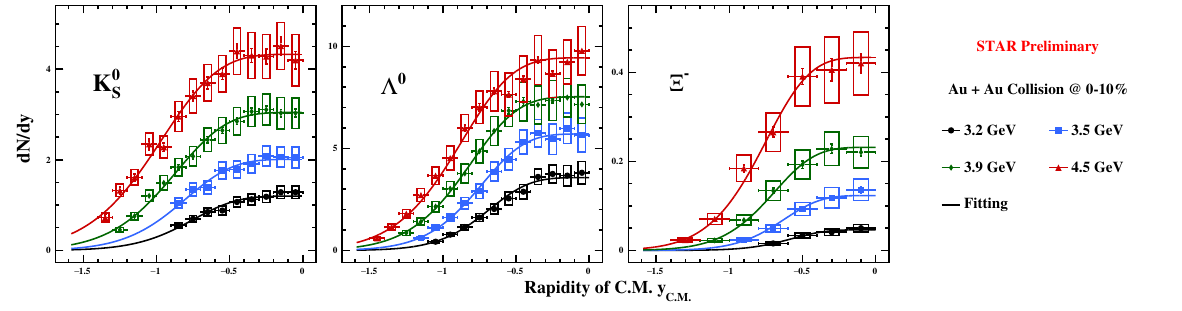}
\vspace{-0.9cm}
\caption{Rapidity distribution of $\rm{K}^0_S, \Lambda$ and $\Xi^-$ in Au+Au central collisions (0-10\%) at $\sqrt{s_{\rm{NN}}}$ = 3.2, 3.5, 3.9 and 4.5 GeV. The vertical line is the statistical uncertainties, and the box is the systematic uncertainties.}
\vspace{-0.8cm}
\label{Fig-2}
\end{figure}

Figure.\ref{Fig-2} shows the rapidity distributions ($\rm{dN}/d\it{y}$) of $\rm{K}^0_S, \Lambda$ and $\Xi^-$ in central (0-10\%) Au+Au collisions. They are obtained by integrating the $p_{\rm T}$ spectra with measured data points and 
fitting extrapolation function. We observe that the strange hadron yields increase with collision energy. The $\rm{dN}/d\it{y}$ shape plateaus at mid-rapidity and is gaussian-like at backward rapidity. In order to describe the $\rm{dN}/d\it{y}$ shape, $\rm{dN}/d\it{y} \propto \frac{\rm 1}{Cosh(y^2/\sigma^2)}$ is chosen and used for fitting data points

\vspace{-0.3cm}

\subsection{Strangeness excitation function}
\label{sec-3-2}

\begin{figure}[h]
\vspace{-0.3cm}
\centering
\includegraphics[width=6.5cm,clip]{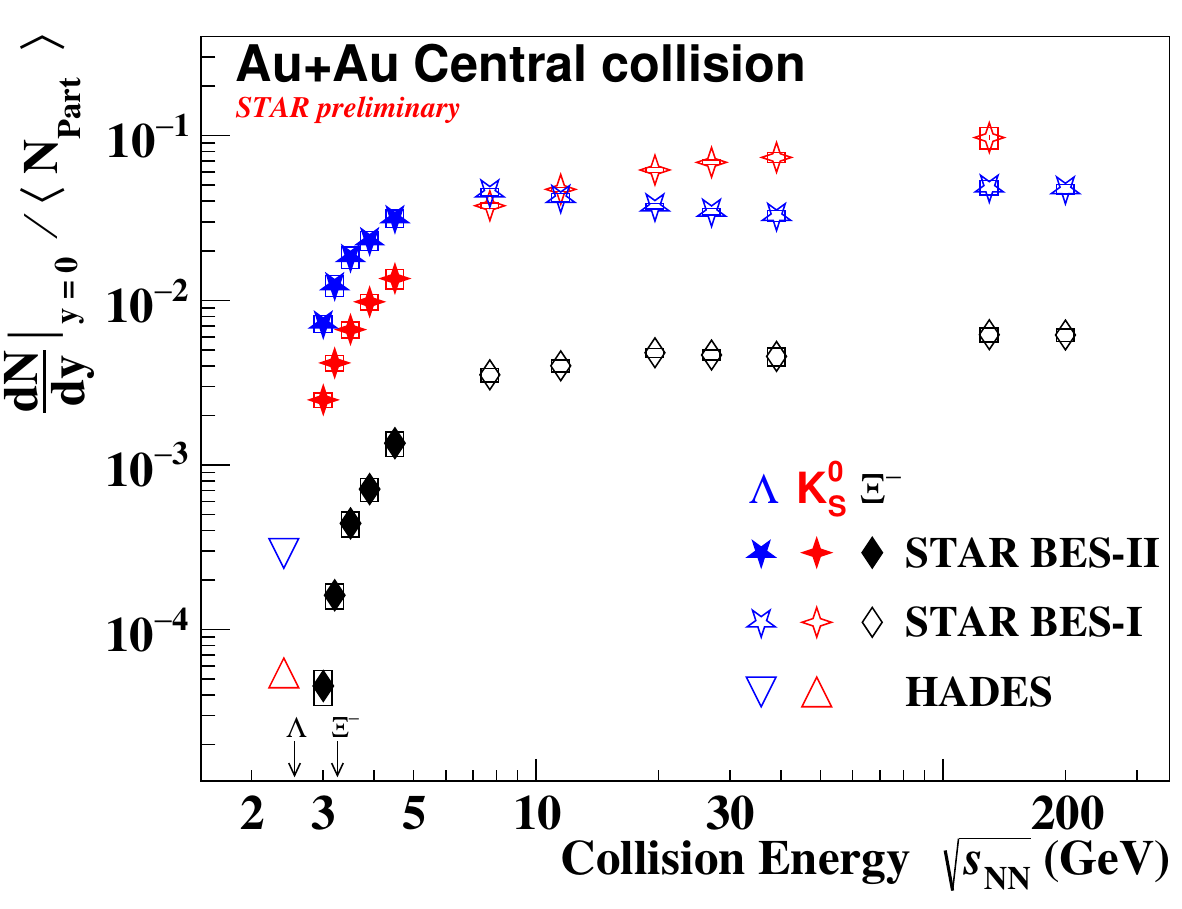}
\vspace{-0.5cm}
\caption{Excitation function of $\rm{K}^0_S, \Lambda$ and $\Xi^-$ in Au+Au the most central collision. The solid points is STAR BES-II FXT-mode results and the open points is STAR BES-I and HADES results\cite{ref-4,ref-5,ref-6,ref-7}. The NN collision threshold energies of $\Lambda$ and $\Xi^-$ are marked by two arrows.}
\label{Fig-4}
\end{figure}

~~~~Excitation functions are obtained from the aforementioned measurements, which is middle rapidity yield of per average participating nucleon number $\langle \rm{N_{part}} \rangle$ as a function of collision energy, and $\langle \rm{N_{part}} \rangle$ is estimated by Glauber model. Figure.\ref{Fig-4} shows the excitation function in Au+Au central collisions.

The STAR-FXT energies crossed NN collision threshold energy of $\Xi^-$ production, so we can see the $\Xi^-$ yield increases rapidly near the threshold. The rate of increase decreases as the collision energy moves away from the threshold, and approximately remains constant at $\sqrt{s_{\rm{NN}}} \sim$ 20 GeV or higher. At low energy $\rm{K}^0_S$ yield are blow the $\Lambda$ ones but that cross at \mbox{$\sim$ 8 GeV} indicating a transition from the baryon-dominated to the meson-dominated matter. The enhancement in baryon yield relative to meson at low energies is likely driven by stronger baryon stopping in the few-GeV energy region.

\vspace{-1.1cm}

\subsection{Scaling property of centrality dependence}
\label{sec-3-3}

\vspace{-0.1cm}

~~~~In order to quantitatively describe the centrality of heavy-ion collisions, we select $\langle \rm{N_{part}} \rangle$ to represent centrality of collision. For determining the centrality dependence of the yields, we fit a power law function and extract a scaling parameter $\alpha_{\rm S}$.

\begin{figure}[h]
\vspace{-0.3cm}
\centering
\includegraphics[width=6.5cm,clip]{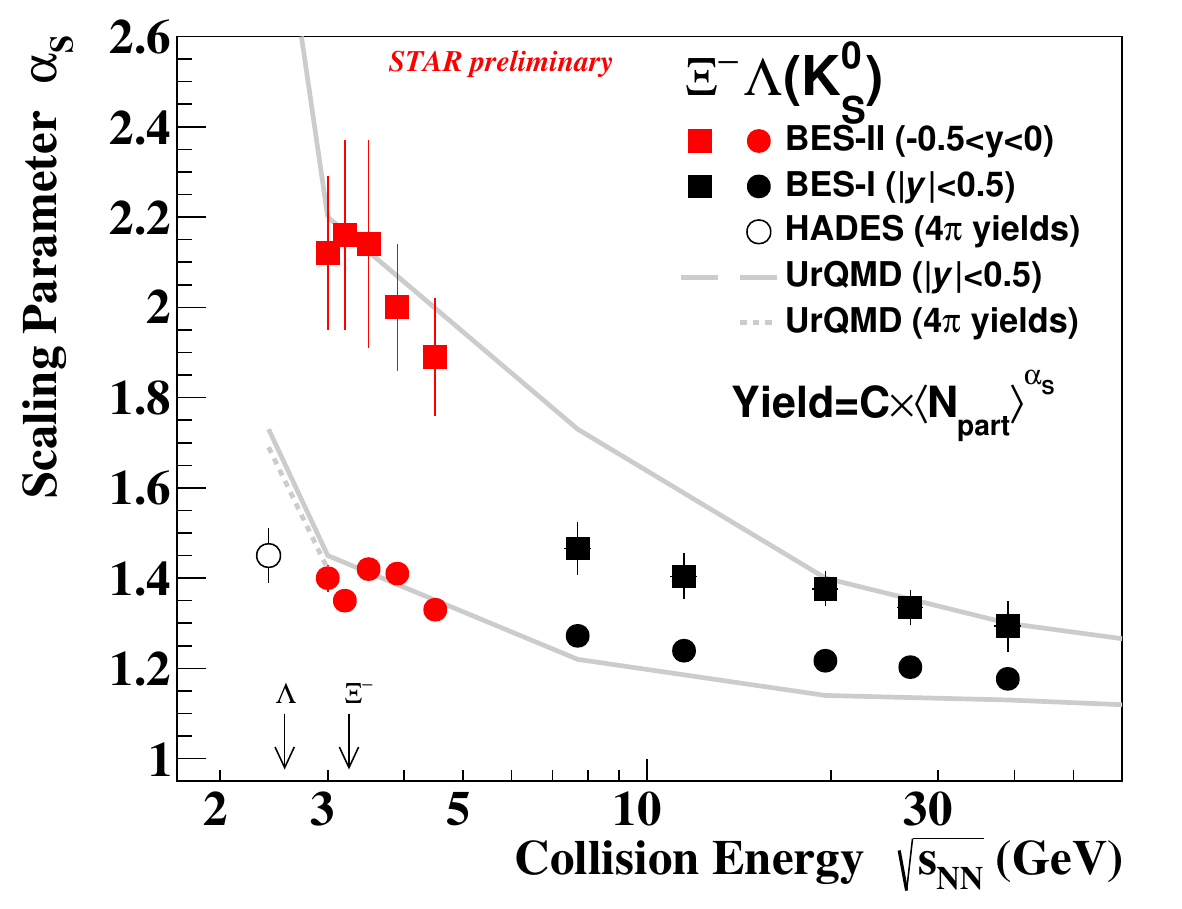}
\vspace{-0.5cm}
\caption{Scaling parameter $\alpha_{\rm S}$ of $\rm{K}^0_S$, $\Lambda$ and $\Xi^-$ as a function of collision energy. Results from the transport model UrQMD is show as gray lines.}
\label{Fig-5}
\end{figure}

Figure.\ref{Fig-5} shows the scaling parameter of $\rm{K}^0_S$, $\Lambda$ and $\Xi^-$ as a function of collision energy. Single strange hadron production $\rm{K}^0_S$ and $\Lambda$ is associated production via $\rm{NN}\rightarrow\rm{N}\Lambda\rm{K}$, therefore their scaling parameter is extracted simultaneously. The scaling parameter is greater than unity, which indicates that their yield increase more rapidly than the increase of the number of participating nucleons. Double strange hadron production $\Xi^-$ has larger scaling parameter than $\rm{K}^0_S$ and $\Lambda$, which indicates that $\Xi^-$ yield increase more rapidly. This may be because $\Xi^-$ has different production channel than $\rm{NN}\rightarrow\rm{N}\Xi\rm{K}\rm{K}$. The scaling parameter decreases with increasing collision energy which shows the centrality dependence of strange hadron yield weakening as collision energy increases.

We also compare this result with transport model UrQMD\cite{ref-8} which qualitatively reproduces the energy dependence, but it cannot quantitatively describe in all energies, especially $\sqrt{s_{\rm{NN}}}$ = 7.7 to 11.5 GeV for $\Xi^-$, which may be due to missing medium effects.

\vspace{-0.1cm}

\section{Summary}
\label{Sec-4}
~~~~In these proceeding, we report yield measurements of $\rm{K}^0_S$, $\Lambda$ and $\Xi^-$ in Au+Au collisions at $\sqrt{s_{\rm{NN}}}$ = 3.2, 3.5, 3.9 and 4.5 GeV. Their $\rm{dN}/d\it{y}$ and the corresponding strangeness excitation function are presented. The decreasing  $\alpha_{\rm S}$ represents centrality dependence of yield as energy increase is shown.


\vspace{0.2cm}

\begin{acknowledgement}
\textbf{Acknowledgement}: This work was supported in part by the National Natural Science Foundation of China (Grant No. 12375134 and No. 12305146), the National Key Research and Development Program of China under Grant No. 2020YFE0202002, and the Fundamental Research Funds for the Central Universities (Grant No. CCNU22QN005).
\end{acknowledgement}




\end{document}